\documentclass[letter,aps,prl,twocolumn,superscriptaddress,showkeys]{revtex4-1}
\usepackage{graphicx}
\usepackage{xcolor}
\usepackage{amsmath, amsfonts, amssymb}
\DeclareGraphicsExtensions{.jpg,.png,.eps,.svg,.pdf}
\usepackage{epstopdf}
\usepackage{subfigure}
\usepackage{enumerate}

\makeindex

\begin{document}

\title{Edge Configurational Effect on Band Gaps in Graphene Nanoribbons}
\author{Deepika}
\affiliation{Department of Physics, Indian Institute of Technology Ropar, Rupnagar-140001, India}
\author{T.J. Dhilip Kumar}
\affiliation{Department of Chemistry, Indian Institute of Technology Ropar, Rupnagar-140001, India}
\author{Alok Shukla}
\affiliation{Department of Physics, Indian Institute of Technology Bombay, Powai, Mumbai-400076, India}
\author{Rakesh Kumar}
\email{Electronic address: rakesh@iitrpr.ac.in}
\affiliation{Department of Physics, Indian Institute of Technology Ropar, Rupnagar-140001, India}
\date{\today}

\begin{abstract}
In this Letter, we put forward a resolution to the prolonged ambiguity in energy band gaps between theory and experiments of fabricated graphene nanoribbons (GNRs). Band structure calculations using density functional theory are performed on oxygen passivated GNRs supercells of customized edge configurations without disturbing the inherent \textit{sp$^2$} hybridization of carbon atoms. Direct band gaps are observed for both zigzag and armchair GNRs, consistent with the experimental reports. In addition, band gap values of GNRs scattered about an average value curve for a given crystallographic orientation are correlated with their width on basis of the edge configurations elucidates the band gaps in fabricated GNRs. We conclude that edge configurations of GNRs significantly contribute to band gap formation in addition to its width for a given crystallographic orientation, and would play a crucial role in band gap engineering of GNRs for future research works on fabrication of nanoelectronic devices.\\

\-\hspace{7.5cm}\textbf{PACS} numbers: 73.22.-f, 73.63.-b, 71.20.-b.
\end{abstract}

\maketitle

Graphene, a single layer of carbon atoms arranged in a honeycomb lattice \citep{novoselov_two-dimensional_2005, geim_rise_2007} is the most interesting material due to its astonishing physical properties like the highest mobility of the charge carriers \citep{bolotin_temperature-dependent_2008}, the highest thermal conductivity \citep{balandin_superior_2008}, and the highest optical transparency at room temperature \citep{nair_fine_2008}. It is a zero band gap material with linear energy-momentum dispersion relation at high symmetry K-points near the Fermi level \citep{wallace_band_1947}, which distinguishes graphene from other two dimensional (2D) layered materials such as \textit{h}-BN, NbSe$_2$, and MoS$_2$ \citep{nagashima_electronic_1995, lebegue_electronic_2009}. Relativistic behaviour of charge carriers as a consequence of the linear dispersion relation assimilates for the first time in condensed matter \citep{novoselov_two-dimensional_2005-1}, and is responsible for extraordinary electronic property like room temperature Quantum Hall effect \citep{geim_graphene:_2007, abanin_spin-filtered_2006}. Even after having miraculous physical properties, the zero band gap of graphene is the biggest roadblock for its applications in electronic devices such as graphene field-effect transistors (gFET) \citep{wang_room-temperature_2008, shukla_graphene_2009, balan_anodic_2010, lin_100-ghz_2010}, and bio-chemical sensors \citep{chowdhury_graphene-based_2011, pumera_graphene_2011, tarakeshwar_hydrogen_2009}, therefore a band gap like semiconductor is required in graphene \citep{novoselov_roadmap_2012}. \\
\indent Band gap in graphene is possible by creating either disorder, chemical doping, strain, functionalization at  edges or surface, or confinement of charge carriers \citep{pereira_modeling_2008, biel_anomalous_2009, lu_band_2010, yan_structural_2009, liu_electrostatic_2009}. But considering the realization for a control over reproducibility of the results, quantum confinement of charge carriers is the most effective way for band gap engineering amongst the possible techniques. In particular, lateral confinement of charge carriers in graphene is introduced by fabricating quasi one-dimensional graphene nanoribbons (GNRs) \citep{nakada_edge_1996}. On the basis of theoretical calculations, it has been acknowledged that in addition to width of GNRs, crystallographic orientation of edges play a key role in the band gap formation \citep{nakada_edge_1996, brey_electronic_2006, son_energy_2006, gundra_theory_2011, gundra_band_2011}. Tight binding calculations show that GNRs of zigzag edges are metallic in nature irrespective of their width, while GNRs of armchair edges are either semiconducting or metallic depending upon their width \citep{nakada_edge_1996}. However, band structure calculations based on density functional theory (DFT) show a small band gap for zigzag GNRs (ZGNRs) on introducing spin into consideration, while armchair GNRs (AGNRs) show band gap correlations depending upon their width, irrespective of spin contribution \citep{son_energy_2006, miyamoto_first-principles_1999}. In contrast to the theory, (a) considerable band gaps are observed in fabricated GNRs \citep{li_chemically_2008, stampfer_energy_2009, han_energy_2007, chen_graphene_2007} and in addition (b) lacks directional correlations with their width \cite{li_chemically_2008,han_energy_2007}. It implies that a parameter other than width and crystallographic orientation plays an influential role in band gap formation of GNRs.\\ 
\indent The aforementioned band structure calculations of GNRs are reported for hydrogen passivated edges, but the most common technique employed to fabricate GNRs is electron beam lithography followed by oxygen plasma etching process \citep{han_energy_2007, chen_graphene_2007, han_electron_2010, berger_electronic_2006}. Therefore, band gap studies require oxygen passivation at edges to study the fabricated GNRs. The theoretical studies carried out so far for oxygen passivated ZGNRs show metallic behaviour  in sharp contrast to the experimental observations \citep{gunlycke_altering_2007, lee_electronic_2009, ramasubramaniam_electronic_2010, vanin_first-principles_2010}. Several efforts have been made to resolve the ambiguity by proposing different theoretical approaches to elucidate the experimental observations either by introducing edge or surface chemistry \citep{yan_structural_2009, lee_electronic_2009}, Coulomb blockade \citep{sols_coulomb_2007}, or edge disorders \citep{evaldsson_edge-disorder-induced_2008, querlioz_suppression_2008, martin_transport_2009}, but have been insufficient. Therefore, it needs more fundamental approach to understand the band gap engineering of fabricated GNRs. We believe that edge configurations may be a crucial parameter in the band gap formation in addition to width and crystallographic orientation of fabricated GNRs. Therefore, we focus our studies on edge configurations to investigate the band gap formation in fabricated GNRs.

\begin{figure}[!t]
\includegraphics[width=9cm,height=6cm,keepaspectratio]{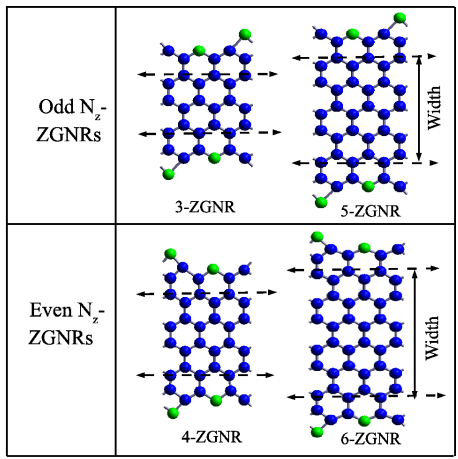}
\caption{(Color online). Edge configurations of oxygen passivated ZGNRs supercells corresponding to odd $N_z$ and even $N_z$. Note a change in the edge configurations of even $N_z$ to odd $N_z$. Width of ZGNRs is shown between the dotted lines. Blue and green spheres represent carbon and oxygen atoms, respectively.}
\label{configzgnr}
\vspace{-5mm}
\end{figure}  

\indent In this Letter, we put forward a resolution to the prolonged ambiguity by proposing a model for the arrangement of oxygen and carbon atoms at the edges of fabricated GNRs. On the basis of DFT calculations, we report (i) direct band gaps for oxygen passivated GNRs consistent with the experimental observations \citep{li_chemically_2008, stampfer_energy_2009, han_energy_2007, chen_graphene_2007}, and (ii) directional correlations in the band gap values are established for fabricated GNRs \citep{li_chemically_2008, han_energy_2007}. 

\begin{figure}[!hb]
\centering
\includegraphics[scale=0.30]{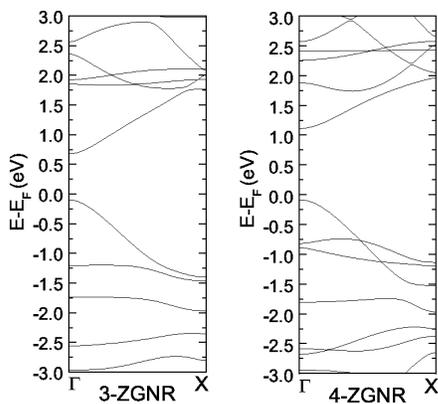}
\caption{Band structure plots from $\Gamma$-point (k = 0) to X-point (k = $\pi$) for ZGNRs of widths corresponding to $N_z$ = 3 and 4. E-E$_F$ is the energy with reference to Fermi level. Note a change in the magnitude of band gap with the edge configuration for even $N_z$ and odd $N_z$.}
\label{bandstzgnr}
\end{figure}

\indent We perform first-principles calculations using DFT as implemented in the Vienna \textit{ab initio} simulation package (VASP) \citep{kresse_efficiency_1996} with generalized gradient approximation (GGA) \citep{perdew_atoms_1992} as exchange-correlation functional. The interactions between electrons-ions are considered using projected augmented wave (PAW) \citep{kresse_ultrasoft_1999} formalism with energy cut-off of 400 eV. The Monkhorst k-space mesh used for GNRs is 25x1x1. All the calculations for GNRs were performed for the nonmagnetics states, assuming doubly occupied bands. A vacuum layer of $\sim10$\AA{} is taken to avoid interactions between the ribbon planes. The GNRs supercells are relaxed until a force on each atom is less than 0.01 eV.\AA{$^{-1}$}. Width of GNRs is represented by number of chains ($N_z$) for ZGNRs as $N_z$-ZGNRs and number of dimer lines ($N_a$) for AGNRs as $N_a$-AGNRs \citep{son_energy_2006}. In our model, oxygen atoms are included in zigzag chains and dimer lines as passivating atoms at edges of ZGNRs and AGNRs, respectively. Periodicity of the GNRs supercells is maintained without disturbing inherent $sp^2$ hybridization of the carbon atoms. The arrangement of oxygen and carbon atoms at the edges of a GNR supercell correspond to an edge configuration, which is different from the edge configurations reported so far for fabricated GNRs \citep{gunlycke_altering_2007, lee_electronic_2009, ramasubramaniam_electronic_2010, vanin_first-principles_2010}.\\
\indent For ZGNRs supercells, the edge configuration changes with its width and results into two configurations corresponding to even $N_z$ and odd $N_z$, respectively [Fig. \ref{configzgnr}]. Width of GNRs is considered up to the outermost chains containing only carbon atoms. We perform band structure calculations for ZGNRs up to a maximum width of 2 nm varying $N_z$ from 1 to 10. Band structures are plotted as a function of k from $\Gamma$-point (k = 0) to X-point (k = $\pi$). Direct band gaps are observed at $\Gamma$-point for all ZGNRs. Typical band structure plots for $N_z$ = 3 and 4 are shown in Fig. \ref{bandstzgnr}. The observation of direct band gaps for all $N_z$-ZGNRs is in sharp contrast to metallic nature of oxygen passivated ZGNRs from the theoretical reports \citep{gunlycke_altering_2007, lee_electronic_2009, ramasubramaniam_electronic_2010, vanin_first-principles_2010}. Projected density of states (pDOS) calculations performed by other groups for oxygen passivated metallic ZGNRs show that $p_z$ orbitals of carbon atoms at the edges of ZGNRs contribute at Fermi level \cite{lee_electronic_2009, ramasubramaniam_electronic_2010}. However, in this work no energy state is observed at the Fermi level for ZGNRs of the proposed edge configurations, and contribution of the $p_z$ orbitals  is shifted below the Fermi level. It implies localization of charges at the edges of ZGNRs due to considerable effect of the edge configurations and nature of the passivating atoms. This localization is obvious in the contour plot of charge density ({\em cf}. Supporting Information), and is the main reason behind the confinement of electrons leading to a finite band gap.

\begin{figure}[!b]
\centering
\includegraphics[scale=0.50]{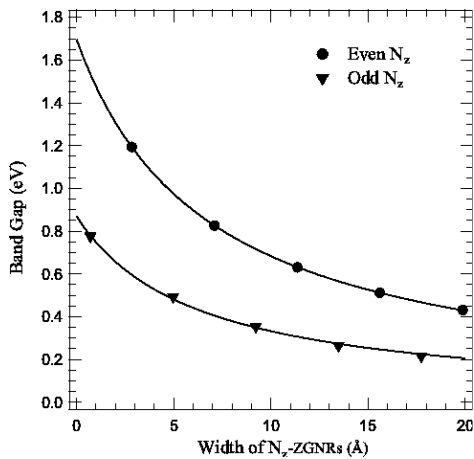}
\caption{Band gap as a function of width for oxygen passivated ZGNRs. Two fitting curves correspond to respective edge configurations of even $N_z$ and odd $N_z$-ZGNRs supercells.}
\label{zgnr}
\end{figure}

\indent Band gap as a function of width for ZGNRs is shown in Fig. \ref{zgnr}. It is observed that the band gap decreases with width, but follows a systematic zigzag pattern. The band gap values are nicely fitted into two curves with a scaling formula given in Eq. (\ref{scalingeqn}) with two sets of fitting parameters. One set of the fitting parameters corresponds to width of even $N_z$ and another set for odd $N_z$, which distinguishes the two fittings as a consequence of the two edge configurations for ZGNRs. However, we note that edge configuration also changes with width for oxygen passivated metallic ZGNRs, but does not produce a band gap \citep{lee_electronic_2009}; similar to hydrogen passivated ZGNRs. Our calculations demonstrate in an explicit manner that more than the width of ZGNRs, it is edge configuration of the oxygen passivated ribbons which causes the opening of band gap. \\
\indent The zigzag pattern of band gaps for ZGNRs are nicely fitted into two curves [Fig. \ref{zgnr}] of following scaling formula,

\begin{equation}
\Delta E = \frac{\alpha}{(w+w^{'})}
\label{scalingeqn}
\end{equation}

\indent Where, $\Delta E$ is band gap (eV), $\alpha$ is scaling factor (eV.\AA), $w$ is width (\AA) of GNRs, and $w^{'}$ is equivalent width (\AA) corresponding to oxygen containing chains or dimer lines. Fitting parameters ($\alpha$, $w^{'}$) are [11.44(1) eV.\AA, 6.74(6) \AA] and [5.39(0) eV.\AA, 6.20(7) \AA] for even $N_z$ and odd $N_z$, respectively. It is to be noted from the experimental reports that band gap values of GNRs plotted against width for a given crystallographic orientation are scattered about an average value curve \citep{han_energy_2007}, which may be a consequence of change in the edge configuration. From the discussions so far, it is concluded that band gap in ZGNRs depends significantly on the edge configuration in addition to its width. 

\begin{figure}[!t]
\centering
\includegraphics[scale=0.30]{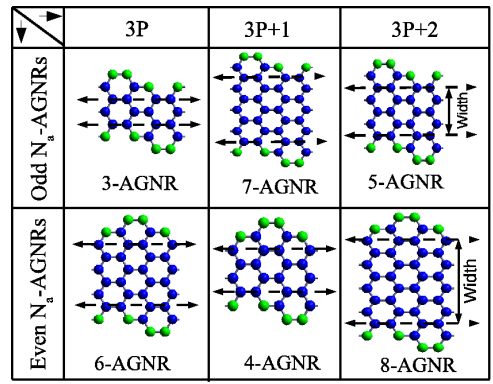}
\caption{(color online) Edge configurations of oxygen passivated AGNRs supercells corresponding to even $N_a$ and odd $N_a$. Note a change in the edge configuration of even $N_a$ to odd $N_a$. Width of AGNRs is considered between the dotted lines. Blue and green spheres represent carbon and oxygen atoms, respectively.}
\label{configagnr}
\end{figure}

\begin{figure}[!t]
\centering
\includegraphics[scale=0.30]{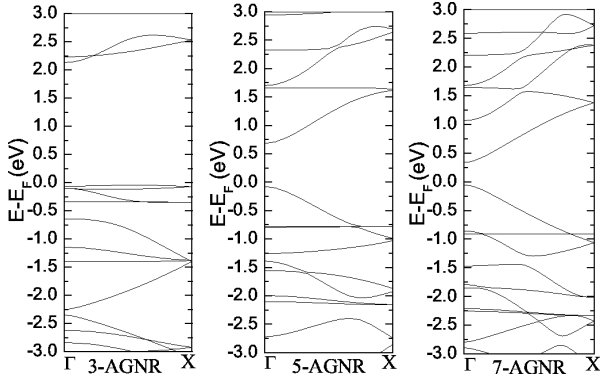}
\caption{Band structure plots from $\Gamma$-point (k = 0) to X-point (k = $\pi$) for AGNRs of widths corresponding to odd $N_a$ = 3, 5, and 7. E-E$_F$ is the energy with reference to the Fermi level.}
\label{bandstagnr}
\end{figure}

\begin{figure}[!t]
\centering
\subfigure[]{\includegraphics[scale=0.55]{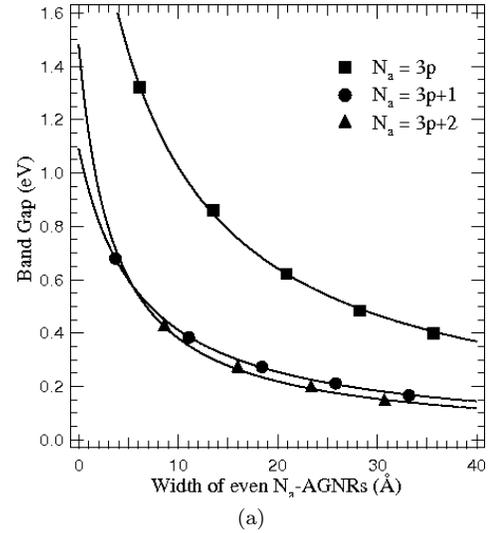}{\label{evenagnr}}}\\
\subfigure[]{\includegraphics[scale= 0.52]{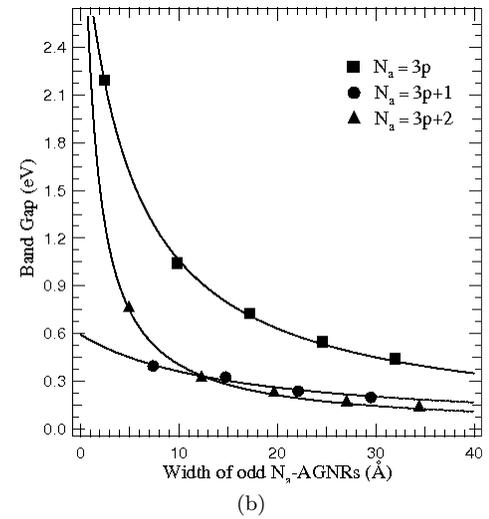}{\label{oddagnr}}}
\caption{Band gap as a function of width for oxygen passivated AGNRs corresponds to respective edge configurations of (a) even $N_a$-AGNRs, (b) odd $N_a$-AGNRs supercells. Band gaps are fitted nicely to three curves corresponding to even $N_a$ and odd $N_a$-AGNRs in 3p, 3p+1, and 3p+2, respectively.}
\label{agnr}
\end{figure}

\indent To explore further the effect of edge configuration, we extend our studies to AGNRs. Edge configuration of AGNRs supercells changes with width and results into two configurations corresponding to even $N_a$ and odd $N_a$ [Fig. \ref{configagnr}], similar to that for ZGNRs. Band structure calculations for AGNRs are performed up to a maximum width of 3.5 nm varying $N_a$ from 1 to 30. Band structures are plotted as a function of k from $\Gamma$-point (k = 0) to X-point (k = $\pi$). Direct band gaps are observed at $\Gamma$-point for all AGNRs. Typical band structure plots for width corresponding to odd $N_a$ = 3, 5, and 7 are shown in Fig. \ref{bandstagnr}; similar band structure plots are observed for even $N_a$-AGNRs. The band gaps are plotted as a function of width for even $N_a$ and odd $N_a$, respectively [Fig. \ref{agnr}]. On fitting, the band gap values are found to be scattered about a curve in the band gap plots for even $N_a$ and odd $N_a$-AGNRs, respectively; similar to the band gap values around a fitting curve for fabricated GNRs \citep{li_chemically_2008, han_energy_2007}. As it is known for hydrogen passivated AGNRs that the band gap values fits into three curves corresponding to width of $N_a$ = 3p, 3p+1, and 3p+2 (where p is a positive integer) \citep{son_energy_2006}. Therefore, we further classify the band gaps corresponding to even $N_a$ and odd $N_a$ into 3p, 3p+1, and 3p+2, respectively; and find nice fitting curves [Fig. \ref{agnr}] with a scaling formula given in Eq. (\ref{scalingeqn}). It is to be noted that without classifying AGNRs into even $N_a$ and odd $N_a$, the band gaps could not be fitted into curves corresponding to width of $N_a$ = 3p, 3p+1, and 3p+2. For even $N_a$-AGNRs, fitting parameters ($\alpha$, $w^{'}$) are [17.23(9) eV.\AA, 6.84(9) \AA],   [6.61(9) eV.\AA, 6.05(7) \AA], and [5.11(6) eV.\AA, 3.44(3) \AA] for width corresponding to 3p, 3p+1, and 3p+2, respectively. For odd $N_a$-AGNRs, fitting parameters ($\alpha$, $w^{'}$) are [15.54(4) eV.\AA, 4.63(9) \AA],   [9.16(5) eV.\AA, 15.43(1) \AA], and [4.41(8) eV.\AA, 0.92(8) \AA] for width corresponding to 3p, 3p+1, and 3p+2, respectively. The values of scaling factor $\alpha$ are consistent with the reported range 2-15 eV.\AA, obtained from DFT calculations of hydrogen passivated AGNRs \citep{son_energy_2006, barone_electronic_2006}. It implies that width dominates over nature of the passivating atoms in band gap formation for AGNRs.

\indent The band gap hierarchy for oxygen passivated AGNRs is found to follow $\Delta E_{3p}$   $>$ $\Delta E_{3p+1}$ $>$ $\Delta E_{3p+2}$, which is different from hierarchy of $\Delta E_{3p+1}$ $>$ $\Delta E_{3p}$ $>$ $\Delta E_{3p+2}$ for hydrogen passivated AGNRs \citep{son_energy_2006, wakabayashi_electronic_1999, yang_quasiparticle_2007}. The change in order may be due to the edge configurations considered for AGNRs supercells. In addition, a crossover is observed for the curves corresponding to width of 3p+1 and 3p+2 for both even $N_a$ and odd $N_a$-AGNRs. Crossover of the curves is observed for width below 15 \AA{} [Fig. 6(a), 6(b)]. Since the edge configurations are same for even $N_a$ and odd $N_a$, respectively; therefore contribution of the edges is expected to be same. But a crossover implies that contribution of oxygen atoms is not same for width belonging to $N_a$ = 3p, 3p+1, and 3p+2 families. Similar to ZGNRs, we envision that edge configuration and width contribute significantly to the band gap formation in AGNRs. 

\indent For a comparative analysis of our calculations with the experimental reports \citep{li_chemically_2008, stampfer_energy_2009, han_energy_2007, chen_graphene_2007}, band gaps of GNRs are estimated using the fitting parameters at width comparable to that of fabricated GNRs. Band gaps are compared for a typical width of 30 nm. For ZGNRs, the estimated band gaps are 0.03(7) eV and 0.01(7) eV for even $N_z$ and odd $N_z$, respectively; which corresponds to an average value of 0.02(7) eV. For even $N_a$-AGNRs, band gaps are 0.05(6) eV, 0.02(1) eV, and 0.01(6) eV for width corresponding to 3p, 3p+1, and 3p+2, respectively. Similarly for odd $N_a$, band gaps are 0.05(1) eV, 0.03(0) eV, and 0.01(5) eV for 3p, 3p+1, and 3p+2, respectively. It corresponds to an average value of 0.03(1) eV for AGNRs. The estimated average band gap for GNRs at 30 nm is 0.02(9) eV, while for fabricated GNRs is 0.01(4) eV \citep{han_energy_2007}. Similar difference in band gap values is observed for other width also. The lower value of band gaps reported for fabricated GNRs may be due to substrate-GNRs interactions \citep{stampfer_energy_2009, han_energy_2007, zhang_electronic_2009}, and overestimation of width \citep{han_energy_2007}.

\indent On the basis of theoretical analysis for both ZGNRs and AGNRs, it is briefed
\begin{enumerate}[(i)]
\item Observation of direct band gaps for both ZGNRs and AGNRs, consistent with the experimental reports \citep{wang_room-temperature_2008, stampfer_energy_2009, han_energy_2007} validates our fundamental approach for studies of fabricated GNRs.
\item The average value of band gaps for ZGNRs and AGNRs are close to each other, and comparable to the experimental reports \citep{han_energy_2007}.
\item The band gaps as a function of width for a given crystallographic orientation are scattered around a fitting curve for GNRs supercells due to change in the edge configuration elucidates the experimental observations \citep{han_energy_2007}.
\end{enumerate}

\indent In conclusion, a resolution for the prolonged ambiguity of band gaps between theory and  experiments of GNRs is presented on the basis of edge configurations. The edge configuration of GNRs significantly contributes to the band gap formation in addition to its width for a given crystallographic orientation. We finally remark that crucial parameters responsible for tuning of a band gap in GNRs are (a) width, (b) crystallographic orientation, (c) edge configuration, and (d) nature of passivating atoms at edges. These insights may greatly help in fine tuning of band gaps for future research works on fabrication of sub-nanometer electronic devices. \\
\indent We gratefully acknowledge CDAC Pune for providing PARAM YUVA II supercomputing facility. One of the authors, Deepika thanks Naresh Alaal and Jagpreet Singh for helpful discussions. 

%

\end{document}